\documentclass[%
 reprint,superscriptaddress,amsmath,amssymb,
 aps,nofootinbib]{revtex4-1}
\usepackage{dsfont}
\usepackage{graphicx}

\usepackage{dcolumn}
\usepackage{bm}
\usepackage{color}
\usepackage{xcolor}
\usepackage{epsfig}
\usepackage{soul}
\usepackage[colorlinks=true,urlcolor=brown,linkcolor=blue,citecolor=magenta]{hyperref}
\usepackage{natbib}
\usepackage{float}
\usepackage{footmisc}
\usepackage{siunitx}
\usepackage[normalem]{ulem}
\usepackage{subcaption}

\definecolor{lightred}{rgb}{1,0.5,0.5}
\definecolor{lightgreen}{rgb}{0.5,1,0.5}
\definecolor{lightblue}{rgb}{0.5,0.5,1}
\definecolor{lightcyan}{rgb}{0.5,0.75,0.75}
\definecolor{lightmagenta}{rgb}{0.75,0.5,0.75}
\definecolor{customgreen}{rgb}{0.494,1,0.502}

\newcommand{\htb}[1]{{\color{black} #1}}
\newcommand{\htm}[1]{{\color{black} #1}}

\begin{document}

\title{\htb{Can WIMPs Survive the Legacy of a Magnetised Early Universe?}}

\author{María Olalla Olea-Romacho }
\email{maria\_olalla.olea\_romacho@kcl.ac.uk}

\author{Malcolm Fairbairn}
\email{malcolm.fairbairn@kcl.ac.uk}

\affiliation{Theoretical Particle Physics and Cosmology, King’s College London, Strand, London WC2R 2LS, United Kingdom 
}
\author{Pranjal Ralegankar}
\email{pralegan@sissa.it}
\affiliation{SISSA - International School for Advanced Studies, Via Bonomea 265, 34136 Trieste, Italy}
\affiliation{INFN – National Institute for Nuclear Physics, Via Valerio 2, I-34127 Trieste, Italy}

\begin{abstract}
Primordial magnetic fields (PMFs) can seed additional small-scale matter fluctuations, leading to the formation of dense, early-collapsing dark matter structures known as minihalos. 
These minihalos may dramatically amplify the dark matter annihilation signal if dark matter is composed of self-annihilating thermal relic particles such as WIMPs. 
In this work, we analyse the annihilation signal from minihalos with prompt central cusps, $\rho \propto r^{-3/2}$, formed due to the enhanced power spectrum induced by PMFs, using gamma-ray observations of the Virgo cluster. We consider benchmarks motivated by cosmological phase transitions, focusing in particular on the electroweak and QCD transitions, where we assume maximal magnetic energy density and horizon-sized coherence length at generation (upper-limit scenarios). In addition, we include a data-driven case corresponding to the best-fit present-day PMF amplitude inferred from DESI BAO and Planck CMB measurements. Under these assumptions, we find that PMFs can place stringent bounds on WIMP annihilation. Magnetic fields with amplitudes matching the DESI–Planck best-fit values are in strong tension with self-annihilating WIMPs across a wide mass range extending beyond the \(\mathrm{TeV}\) scale, while the electroweak- and QCD-phase-transition toy-model benchmarks would exclude thermal relics with masses below $300\,\mathrm{GeV}$ and $3\,\mathrm{TeV}$, respectively. Although weaker PMFs would yield weaker annihilation signals, our results demonstrate that whenever PMFs enhance small-scale structure, indirect-detection limits on dark matter must be revisited.
\end{abstract}

\maketitle

\section{Introduction}

Magnetic fields are observed on a wide range of astrophysical scales, from galaxies to galaxy clusters and the intergalactic medium \cite{doi:10.1126/science.1184192, HESS:2014kkl, Finke:2015ona, VERITAS:2017gkr, AlvesBatista:2021sln, MAGIC:2022piy, Vovk:2023qfk, Tjemsland:2023hmj,Broderick:2011av}. Their origin, however, remains an open problem. One compelling possibility is that they were seeded by processes in the early universe, such as inflationary dynamics \cite{Ratra:1991bn, Turner:1987bw, Kobayashi:2014sga, Durrer:2010mq, Sharma:2018kgs, Yanagihara:2023qvx} 
or cosmological phase transitions~\cite{Vachaspati:1991nm, Ellis:2019tjf, Olea-Romacho:2023rhh,Quashnock:1988vs,Sigl:1996dm, Yang:2021uid, Di:2020kbw}. If primordial magnetic fields (PMFs) exist, they may leave observable signatures on cosmological and astrophysical observables, including the cosmic microwave background~\cite{Trivedi:2018ejz, Paoletti:2022gsn, Wagstaff:2015jaa, Jedamzik:1999bm,Planck:2015zrl, Paoletti:2019pdi,LiteBIRD:2024twk}, structure formation~\cite{wasserman97, Kim:1994zh, Subramanian:1997gi,Shaw:2010ea, Fedeli2012,pandeysethi:2013,Chong13,khan13,montanino2017,Sanati:2020,Sanati:2024ijt,Zhang:2024yph,Katz:2021, Tashiro:2006, Sethi:2004pe, pandeysethi:2015,Cruz:2023rmo,Bhaumik:2024efz,Adi:2023doe, Schleicher:2008hc,Ralegankar:2024ekl, Ralegankar:2024arh,Pavicevic:2025gqi}, and gravitational wave backgrounds~\cite{RoperPol:2022iel, RoperPol:2022hxn, Balaji:2024rvo}. Recently, PMFs have also been invoked to alleviate the Hubble tension between values of $H_0$ obtained through supernova observations and through observations of the CMB \cite{Riess:2020fzl, Planck:2018vyg,Jedamzik:2025cax, Jedamzik:2020krr}. 

\begin{figure}[h!]
    \centering
    \includegraphics[width=\columnwidth]{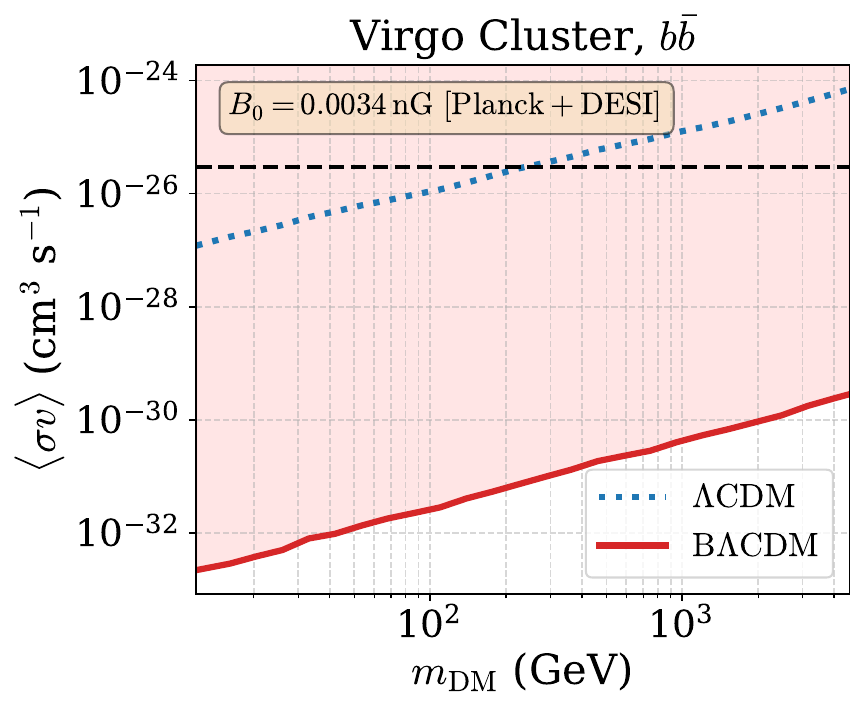}
    \caption{Upper limits on the dark matter annihilation cross-section to a $b \bar{b}$ final state in the Virgo cluster, considering a primordial magnetic field with present strength $B_0 = 3.4 \, \mathrm{pG}$, corresponding to the best fit value from the
    combination of Planck CMB data and the baryon acoustic oscillation (BAO) measurements from the DESI Year 1 release~\cite{Jedamzik:2025cax}. Our limits rule out
    the thermal annihilation cross-section (black dashed) for dark matter masses extending beyond the $\mathrm{TeV}$ scale. \htb{The curve labelled ``$\Lambda$CDM'' corresponds to $\Lambda$CDM cosmology with self-anihilating dark matter but without any enhancement from primordial magnetic fields, while the B$\Lambda$CDM curve include the additional prompt cusps seeded by PMFs. For the assumptions on the evolution of the magnetic field strength and coherence length, see Sec.~\ref{sec:PMFandHaloFor}.}}
    \label{fig:virgoPlanck}
\end{figure}

A consequence of PMFs is their ability to enhance the growth of small-scale matter perturbations~\cite{Ralegankar:2023pyx}. Prior to recombination, the Lorentz force from a stochastic magnetic field induces compressible modes in the baryon fluid on scales smaller than the photon mean free path. While these baryonic perturbations can subsequently be erased by baryon thermal pressure, they leave a persistent gravitational imprint on the dark matter component. As a result, the dark matter power spectrum is amplified on small scales, potentially giving rise to a population of early-forming, compact minihalos \cite{Ralegankar:2023pyx}.

In parallel, thermal relic dark matter candidates, such as weakly interacting massive particles (WIMPs), remain a key target of experimental efforts~\cite{LZ:2024zvo}. These particles are expected to self-annihilate, producing observable Standard Model final states, including gamma rays. However, in the standard cold dark matter (CDM) scenario, the annihilation signal from small-scale halos is typically suppressed by the late formation time and relatively shallow central profiles of low-mass halos.

This suppression can be overcome in scenarios with PMFs. The early formation of dark matter minihalos from sharp peaks in the enhanced power spectrum leads to the formation of \emph{prompt cusps}~\cite{Delos:2019dyh}: dense, centrally-concentrated structures with $\rho \propto r^{-3/2}$. \htb{These objects are theoretically motivated as the first gravitationally bound dark-matter structures to form in the Universe, since the dark-matter free-streaming length suppresses the growth of smaller perturbations and prevents hierarchical merging at earlier times. 
Prompt cusps therefore provide a natural minimal building block in standard $\Lambda$CDM. In the presence of primordial magnetic fields this assumption is particularly well–motivated, since the enhanced small–scale power produced by PMFs determines the first scales to become non–linear and collapse, naturally seeding such cusps. If, in addition, the dark matter is self-annihilating, they significantly enhance the annihilation rate and the expected gamma-ray signal, especially in environments such as galaxy clusters
}~\cite{Crnogorcevic:2025nwp, Delos:2022bhp}.

In this work, we investigate the indirect detection prospects of self-annihilating dark matter in the presence of PMFs. Specifically, we analyze how PMF-induced prompt cusps modify the annihilation signal from the Virgo cluster, and use this to place constraints on the dark matter annihilation cross section. We consider three benchmark scenarios for the magnetic field strength: (i) a field generated at the QCD phase transition, (ii) a weaker field generated at the electroweak phase transition, and (iii) the best-fit field amplitude obtained from a joint analysis of Planck and DESI data~\cite{Jedamzik:2025cax} (see Fig.~\ref{fig:virgoPlanck}).

Generically, we find that the magnetic fields in the early universe places stringent constraints on the WIMP annihilation cross section. In particular, PMFs generated both around the QCD phase transition and also with cosmologically-motivated values exclude thermal WIMPs over a wide mass range extending beyond the $\mathrm{TeV}$ scale, potentially challenging their viability as dark matter candidates.  For magnetic fields generated at the electroweak phase transition, we exclude thermal-relic WIMPs with masses \( m_{\chi} \lesssim 300 \, \mathrm{GeV} \), with constraints only marginally stronger than those obtained in \(\Lambda\)CDM \htb{with self-annihilating dark matter but no primordial magnetic fields}~\cite{Crnogorcevic:2025nwp}.
 More generally, magnetic fields produced at phase transition temperatures \( T \lesssim 100 \, \mathrm{GeV} \) can exclude WIMPs with masses below \( 400 \, \mathrm{GeV} \), further tightening the bounds on the low-mass window.
This work demonstrates that gamma-ray searches can be used to probe not only the properties of dark matter, but also the magnetic history of the early universe.

\section{Primordial magnetic fields and minihalo formation}
\label{sec:PMFandHaloFor}

Primordial magnetic fields can imprint significant features on the small-scale matter power spectrum~\cite{Ralegankar:2023pyx}. A stochastic primordial magnetic field exerts a compressive Lorentz force on the baryon fluid. While baryons remain tightly coupled to the photon bath, relativistic photon pressure suppresses any growth on scales larger than the photon mean free path. However, below this scale, the coupling weakens, allowing the Lorentz force to drive baryonic inhomogeneities.

These baryonic overdensities may ultimately be erased by baryon thermal pressure. However, during the time in which the baryon fluctuations grow, they source gravitational potentials that also affect the dark matter component. Since dark matter is collisionless and does not couple directly to photons or magnetic fields, it remains unaffected by baryon thermal pressure and retains the memory of these early gravitational perturbations. As a result, the dark matter power spectrum is enhanced on small scales, even below the baryon Jeans length. The growth of the dark matter component continues uninterrupted after recombination, leading to the formation of compact, high-density dark matter structures at early times, commonly referred to as minihalos.

\htb{In this study we remain agnostic about the precise magnetogenesis mechanism. 
Our analysis applies equally to fields generated at specific phase transitions 
(such as the QCD or electroweak transitions) or to more generic stochastic fields characterised by a present-day strength $B_{0}$ and coherence length $\xi_{0}$.  
We assume that the field is initially non-helical, evolves after generation according to standard magnetohydrodynamic processes, and has a Batchelor spectrum on large scales, $P_B(k)\propto k^2$.

To model the field evolution we assume reconnection–driven turbulence when the plasma kinetic Reynolds number is large ($Re>1$), and viscous damping once $Re<1$. 
While $Re>1$, the field undergoes an inverse cascade conserving the Hosking invariant~\cite{Hosking:2020wom,Hosking:2022umv},
\begin{equation}
B^4\,\xi^5 = \mathrm{const}\,.
\end{equation}
In this turbulent regime the coherence scale grows with time and is roughly given by~\cite{Hosking:2022umv},\footnote{Note that the prefactor in eq.~\eqref{eq:xi_turb} can depart from 0.1 depending on the magnetic Prandtl number and the Lundquist number \cite{Hosking:2022umv}. In regimes of our interest, i.e. where PMFs influence dark matter perturbations, we find that the prefactor is $\sim 0.1$ \cite{Ralegankar:2023pyx}.}
\begin{equation}\label{eq:xi_turb}
\xi(a) \;\sim\; \frac{0.1\,V_A}{a\,H(a)}\,,
\end{equation}
where $H(a)$ is the Hubble parameter at redshift $a$ and $V_A$ is the Alfven speed defined as 
\begin{align}
    V_A=\frac{B}{\sqrt{4\pi\rho_{\rm plasma}}}.
\end{align}
Here $\rho_{\rm plasma}$ is the energy density of particles that are tightly coupled to the baryon plasma. For instance, on scales smaller than the photon or neutrino diffusion lengths, $\rho_{\rm plasma}$ does not include photons/neutrinos. 

The turbulence in the PMF-driven plasma is suppressed if the neutrino or photon diffusion scale ($\ell_{D}$) exceeds the PMF coherence scale while remaining smaller than the horizon size, i.e $\xi<\ell_{D}<(aH)^{-1}$. In this regime, interactions between baryons and neutrinos or photons act as an effective viscosity, with the Reynolds number governed by $Re \sim V_A \xi / \ell_{\rm mfp}$, where $\ell_{\rm mfp}$ denotes the relevant mean free path (photon/neutrino). Using eq.~\eqref{eq:xi_turb}, one can show that $Re<1$ is equivalent to photon/neutrino diffusion length scale exceeding PMF coherence scale. Once this occurs, the coherence scale follows the approximate relation~~\cite{Ralegankar:2023pyx,Banerjee:2004df} 
\begin{equation}\label{eq:xi_D_visc}
\xi(a) \;\simeq\; \frac{V_A}{a\,\sqrt{(\ell_{\rm mfp}^{-1}+H(a))H(a)}}\,.
\end{equation}
Furthermore, in this viscous regime, we assume that the Hosking integral is no longer conserved and that small-scale PMFs decay without any inverse-cascade. For the Batchelor–spectrum case considered here, this leads to the scaling relation~\cite{Banerjee:2004df} 
\begin{equation}\label{eq:alfven_cons}
B^2\,\xi^5 = \mathrm{const}\, .
\end{equation}
We note that this assumption differs from more recent treatments, which propose that the Hosking invariant remains conserved even in viscous regimes~\cite{Hosking:2022umv}. However, no dedicated simulations have yet confirmed that conservation, while analytical studies support the validity of eq.~\eqref{eq:alfven_cons} in this regime~\cite{Ralegankar:2023pyx,Campanelli:2013iaa}.
The viscous–drag phase ends once the corresponding mean free path (neutrino or photon) exceeds the horizon size, marking the onset of free–streaming.
It is worth noting that the indirect–detection constraints from dark–matter annihilation are not directly sensitive to the full early–Universe evolution of primordial magnetic fields. They become relevant only once PMFs begin to affect the dark–matter density perturbations—that is, after the baryon plasma becomes compressible, when photons decouple from the magnetized baryon fluid. Consequently, the results presented here are primarily sensitive to the PMF evolution after the start of the photon–drag (viscous) regime.

We define $a_I$ as the epoch, when $Re$ falls below unity due to photon viscosity,
\begin{equation}
k_I(a_I)=\xi^{-1}(a_I)=\ell_{\gamma D}^{-1}(a_I)\,.
\end{equation}
For $a<a_I$, we evolve PMFs with approximate power-laws mentioned above. \htm{For $a > a_I$, we explicitly solve the evolution of PMFs along with their impact on baryon and dark matter perturbation equations following the prescription in~\cite{Ralegankar:2023pyx}. The comoving magnetic field strength is then updated consistently with this damping evolution, ensuring a smooth transition between the turbulent and viscous regimes.
At late times ($a \ge a_{\rm rec}$), the field is effectively frozen into the plasma, so the comoving field strength and coherence length remain constant \cite{Banerjee:2004df}.
In what follows, we adopt this evolution prescription for $B(a)$ and $\xi(a)$ consistently across all benchmark scenarios analysed in this work, and illustrate it in Sec.~\ref{sec:results}.}}

The abundance and internal structure of minihalos depend sensitively on the amplitude and shape of the enhanced power spectrum induced by the PMFs. In particular, PMFs with a Batchelor spectrum on large scales, $P_B(k) \propto k^2$, can lead to enhancements at sub-solar mass scales, potentially producing minihalos as light as $10^{-11}\,M_\odot$. The Bachelor spectrum is expected from PMFs generated in phase transition. Moreover, recent studies suggest that the subsequent turbulent evolution after phase transition enforces the Batchelor spectrum even if the initial field had a different spectrum \cite{Brandenburg:2023rrd}.

This population of early-forming minihalos provides a novel probe of primordial magnetism, especially in scenarios where dark matter is self-annihilating, which is the subject of this paper. The initial collapse of these small-scale overdensities can lead to the formation of dense central regions known as prompt cusps, which play a key role in enhancing the indirect detection signal.

In this study, we connect the enhancement in the power spectrum to the initial PMF parameters using the analytical framework developed in Ref.~\cite{Ralegankar:2023pyx}. A central assumption of this method is that the PMFs follow a Gaussian distribution. Given the possibility of non-Gaussianities in a realistic scenario, our findings should be regarded as order-of-magnitude estimates. 
\htm{In this approach, the Lorentz force sourced by stochastic PMFs drives baryon and dark matter perturbations through the following Boltzmann equations, 
\begin{align}
&a^2\frac{d^2 \delta_{\rm DM}}{da^2} + 
a\left[1+\frac{d{\rm ln}(a^2H)}{d{\rm ln}a}\right]\frac{d\delta_{\rm DM}}{dy}
- \frac{3}{2}\frac{\Omega_{\rm DM}}{\Omega(a)}\delta_{\rm DM} \nonumber\\
&= \frac{3}{2}\frac{\Omega_b}{\Omega(a)}\delta_b,
\end{align}
\begin{align}\label{eq:deltab_a2}
		&a^2\frac{\partial^2 \delta_{\rm b}}{\partial a^2}+a\left[1+\frac{d{\rm ln}(a^2H)}{d{\rm ln}a}+\frac{4\rho_{\gamma}}{3\rho_b\ell_{\gamma, {\rm mfp}}H}\right]\frac{\partial \delta_{\rm b}}{\partial a}+\frac{k^2c_{\rm b}^2}{(aH)^2} \delta_{\rm b}\nonumber\\ &-\frac{3}{2}\frac{\Omega_{\rm b}}{\Omega(a)}\delta_{\rm b}=-\left[\frac{3M_{\rm Pl}^2S_B}{\rho_{\rm m0}}\right]\frac{\Omega_{\rm m}}{\Omega(a)}+\frac{3}{2}\frac{\Omega_{\rm DM}}{\Omega(a)}\delta_{\rm DM}.
	\end{align}
where $\delta_{\rm DM}$ and $\delta_b$ denote the fractional density perturbations of dark matter and baryons, respectively.
The parameters $\Omega_{\rm DM}$, $\Omega_b$, and $\Omega_m$ are the present-day dark matter, baryon, and total matter density fractions, respectively. We ignore the contribution from dark energy as the halos we are concerned with form much before dark energy becomes important. Thus, we approximate $\Omega(a)= \Omega_{\rm m}(1+a_{\rm eq}/a)$ with $a_{\rm eq}$ being the matter-radiation equality. The source term $S_B$ represents the Lorentz force exerted by PMFs,
\begin{align}\label{eq:S0}
		S_B=\frac{\nabla\cdot[(\nabla\times\vec{B})\times\vec{B}]}{4\pi \rho_{\rm b0}}.
	\end{align}
Integrating the above equations yields the magnetically induced dark–matter perturbation $\delta_{\rm DM,B}$, 
from which we compute the magnetic contribution to the power spectrum, 
$P^B_{\rm DM}(k,z)=\langle |\delta_{\rm DM,B}(k,z)|^2 \rangle$. 
We then define the total power spectrum as
\[
P^{\rm tot}_{\rm DM}(k,z) = P^{\Lambda{\rm CDM}}_{\rm DM}(k,z) + P^B_{\rm DM}(k,z),
\]
which is used as input for the halo–formation model described in Sec.~\ref{sec:PCprop}. 
This procedure reproduces the enhancement of the transfer function $T^2(k)$ 
shown in Fig.~2 of Ref.~\cite{Ralegankar:2023pyx}, 
linking the PMF parameters $(B_0,\xi)$ to the abundance of magnetically induced minihalos. For more technical details, we refer to Ref.~\cite{Ralegankar:2023pyx}.
}

A key physical scale in this context is the free-streaming length of dark matter, which suppresses structure formation above a characteristic wavenumber $k_{\rm fs}$. This suppression arises because dark matter particles decouple kinetically from the plasma at a finite temperature and retain residual velocities that erase small-scale perturbations through free-streaming. The free-streaming wavenumber $k_{\rm fs}$ depends on both the dark matter mass $m_{\chi}$ and the kinetic decoupling temperature $T_{\rm kd}$~\cite{Green:2005fa}:
\begin{align}
k_{\mathrm{fs}} &\approx 1.70 \times 10^6\, \mathrm{Mpc}^{-1} 
\left( \frac{m_{\chi}}{100\,\mathrm{GeV}} \right)^{1/2} \\
&\times \left( \frac{T_{\mathrm{kd}}}{30\,\mathrm{MeV}} \right)^{1/2}
\left[ 1 + \frac{\ln(T_{\mathrm{kd}} / 30\,\mathrm{MeV})}{19.2} \right]^{-1}.
\end{align}
To account for this physical suppression in our calculation, we imposed a smooth cutoff on the matter power spectrum at $k_{\rm fs}$, modeling it with the function
\begin{equation}
f(k) = \frac{1}{1 + \exp\left( \frac{k - k_{\mathrm{\rm fs}}}{\delta_k  k_{\mathrm{\rm fs}}} \right)},
\end{equation}
where the parameter $\delta_k$ controls the smoothness of the cutoff.

In all our results, we fixed the kinetic decoupling temperature at \( T_{\mathrm{kd}} = 30 \, \mathrm{GeV} \). Realistic dark matter models that evade current direct detection constraints often feature early kinetic decoupling. For instance, pseudo Nambu–Goldstone boson dark matter models~\cite{Gross:2017dan, Arcadi:2025sxc}, which naturally suppress tree-level scattering with nuclei, can exhibit kinetic decoupling temperatures comparable to or even exceeding the freeze-out temperature~\cite{Abe:2021jcz}. Moreover, since the annihilation signal depends only logarithmically on \( T_{\mathrm{kd}} \), our results are only mildly sensitive to this parameter: for benchmark magnetic field strengths of order \( \mathrm{pG} \), varying \( T_{\mathrm{kd}} \) from \( 30 \, \mathrm{MeV} \) to \( 30 \, \mathrm{GeV} \) changes the cross-section constraints by at most \( \sim 160\% \) and by as little as \( \sim 12\% \) across the full dark matter mass range. This mild dependence reflects the fact that the properties of prompt cusps are mainly dictated by the scale of the power spectrum enhancement, rather than by the free-streaming cutoff.

\section{Prompt cusps properties}
\label{sec:PCprop}

Prompt cusps are dense, power-law central structures that form during the monolithic (i.e. not hierarchical) collapse of isolated peaks in the early dark matter density field. Unlike the more familiar Navarro-Frenk-White (NFW) halos, which arise from hierarchical structure formation and exhibit a characteristic core-like inner slope (\( \rho \propto r^{-1} \)), prompt cusps develop  an inner density profile of \( \rho \propto r^{-3/2} \)~\cite{Delos:2019mxl, Delos:2022bhp}. This profile forms quasi-instantaneously at the moment of collapse and is tightly determined by the local properties of the linear density peak, particularly its amplitude and curvature. The prompt cusp persists at the center of the resulting halo even as it later accretes additional material and grows an NFW-like outer envelope~\cite{Delos:2022bhp}.

Importantly, while in $\Lambda$CDM cosmologies the smallest halos form at the free-streaming scale $k_{\rm fs}$, in PMF cosmologies it is the scale of the spectral bump that determines the first structures to collapse, as it is the first to become non-linear. This is because the bump injects a localised excess of power at scales larger than the free-streaming length, boosting the density contrast sufficiently for those perturbations to reach the collapse threshold earlier than any others. As a result, the earliest minihalos form around the bump scale rather than the free-streaming cutoff. In particular, the gravitational influence of magnetically induced baryon inhomogeneities seeds dark matter perturbations on these intermediate scales, enhancing the abundance of low-mass minihalos, and increasing the likelihood that they form prompt cusps: dense, centrally concentrated structures with $\rho \propto r^{-3/2}$~\cite{Ralegankar:2023pyx}.

Each prompt cusp forms with a density profile \( \rho = Ar^{-3/2} \) extending between an inner core radius \( r_{\rm core} \), imposed by phase-space constraints, and an outer cusp radius \( r_{\rm cusp} \), determined by the peak size. Since the annihilation rate formally diverges at small radii in a density profile that scales as \( \rho \propto r^{-3/2} \), understanding the inner structure of prompt cusps is essential. Analytical arguments~\cite{Delos:2022yhn}, supported approximately by numerical simulations~\cite{Maccio:2012qf}, indicate that the initial thermal motion of dark matter particles imposes a finite-density core at the center of the cusp. The radius of this core, \( r_{\mathrm{core}} \), is given by 
\begin{equation}
    r_{\mathrm{core}} \simeq 0.34\, G^{-2/3} \left( \frac{m_{\chi}}{T_{\mathrm{kd}}} \right)^{-2/3} \bar{\rho}(a_{\mathrm{kd}})^{-4/9} A^{-2/9},
\end{equation}
where \( \bar{\rho}(a_{\mathrm{kd}}) \) is the mean dark matter density at the time of kinetic decoupling, and \( A \) is the normalisation of the cusp density profile
\begin{equation}
    \rho = A r^{-3/2}.
\end{equation}
 The coefficient \( A \) can be statistically predicted from the linear power spectrum~\cite{Delos:2019mxl}:
\begin{equation}
    A \simeq 24\, \bar{\rho}(a_{\mathrm{coll}})\, (a_{\mathrm{coll}} R)^{3/2}.
\end{equation}
and is set by the collapse time $a_{\mathrm{coll}}$ and  characteristic comoving size $R$.
The scale factor at collapse $a_{\mathrm{coll}}$ is estimated using the ellipsoidal collapse approximation in Ref.~\cite{Sheth:1999su}, and the characteristic comoving size  is given by $R\equiv|\delta/\nabla^2\delta|^{1/2}$, where $\delta$ is the height of the peak in the density contrast and $\nabla^2\delta$ is the Laplacian of the density contrast with respect to the comoving position. 
This cusp profile extends out to a radius  set by the comoving size of the collapsing density peak and its collapse time, and is given by~\cite{Delos:2019mxl}
\begin{equation}
    r_{\mathrm{cusp}} \simeq 0.11\, a_{\mathrm{coll}} R.
\end{equation}

These structures are the densest dark matter configurations expected in a CDM universe and are predicted to survive hierarchical merging and tidal stripping with their inner structure largely intact~\cite{Delos:2022bhp}. Since we will be studying the signal from the halos of clusters of galaxies, rather than within galactic halos themselves, we do not expect significant tidal disruption.

 To model the properties of prompt cusps seeded by enhanced small-scale fluctuations, we make use of the publicly available code developed in~\cite{Delos:2019mxl,Delos:2022yhn}, which samples halo populations from a given matter power spectrum. The code implements the BBKS peak statistics formalism~\cite{Bardeen:1985tr} and uses the ellipsoidal collapse criterion to determine whether a peak collapses to form a prompt cusp. For each collapsing peak, it predicts the asymptotic density profile.
 
\section{Annihilation signal}

The steep inner density profiles of prompt cusps significantly enhance the dark matter annihilation rate relative to standard halo models. The high central densities of these structures result in substantial contributions to the annihilation signal, even though the cusps themselves comprise only a small fraction of the total dark matter mass (around the $5 \%$).

To quantify the contribution of prompt cusps to the annihilation signal, we follow the formalism introduced in~\cite{Delos:2022bhp} and developed in~\cite{Crnogorcevic:2025nwp}. The gamma-ray flux from dark matter annihilation reads
\begin{equation}
    \frac{\rm{d}^2\Phi}{\rm{d} \Omega\, dE} = \frac{\langle \sigma v \rangle}{8\pi m_\chi^2} \frac{\rm{d} N_\gamma}{ \rm{d} E} \frac{\rm{d} \rm J}{\rm{d}\Omega}.
\end{equation}
Here \( \langle \sigma v \rangle \) is the thermally averaged annihilation cross section,  and \( \rm{d} N_\gamma / \rm{d} E \) is the gamma-ray spectrum per annihilation. The astrophysical contribution is encoded in the \( \rm{J} \)-factor, which in our case includes both the smooth halo and the contribution from surviving prompt cusps
\begin{equation}
    \frac{\rm{dJ}}{\rm{d}\Omega} = \int d\ell\, \rho(r) \left[ \rho(r) + f_{\rm surv} f_{\rm tidal}(r) \rho_{\rm eff, 0} \right],
\end{equation}
integrated over the line-of-sight $l$.
Here, \( \rho(r) \) is the smooth density profile of the host halo (for which we adopt an NFW profile), and the second term captures the contribution from embedded prompt cusps. The parameter \( f_{\rm surv} \sim 0.5 \) accounts for the fraction of cusps that survive mergers and tidal disruption after accretion into larger structures~\cite{Delos:2022bhp}, and \( \rho_{\rm eff, 0} \) is the effective annihilation density of the initial population of cusps. The quantity \( \rho_{\rm eff, 0} \) is computed by statistically sampling the cusp population from the enhanced matter power spectrum, following the procedure outlined in the previous section. \htb{The survival fraction $f_{\rm surv}$, which quantifies the proportion of prompt cusps that survive mergers and tidal disruption, 
is taken to be $f_{\rm surv}\simeq0.5\pm0.1$ following Refs.~\cite{Delos:2022bhp}. This value is conservative for our CMB-motivated and QCD-PT benchmarks (see Sec.~\ref{sec:results}), whose more massive minihalos form with lower number density, 
reducing the rate of encounters and mergers. 
For lighter halos such as those formed at the electroweak phase transition, this assumption may need revision, 
but we adopt $f_{\rm surv}=0.5$ here for consistency across benchmarks.
}

The average contribution of the prompt cusps per mass of dark matter is:
\begin{equation}\label{eq:rhoeff0}
     \rho_{\rm eff, 0}
    \equiv \frac{\int_\mathrm{cusps}\rho^2 \rm d V}{M}
    =n_\mathrm{peaks}\langle j\rangle/\bar\rho_0,
\end{equation}
where $n_{\rm peaks}$ is the comoving number density of collapsed dark matter structures that form from individual peaks in the linear density field~\cite{Delos:2019mxl}, and $\rho_0 \simeq 33\, M_\odot\,\mathrm{kpc}^{-3}$.
We sampled $N=10^4$ density peaks, and averaged over the population to obtain $\langle j\rangle$ defined through
\begin{align}\label{eq:Jcusp}
    j \equiv \int_\mathrm{cusp}\rho^2 \rm d V = 4\pi A^2[1/3+\rm {ln}(r_{\rm cusp}/ r_{\rm core})].
\end{align}

In our analysis, we focus on the Virgo cluster as the observational target. This choice is motivated by the results in~\cite{Crnogorcevic:2025nwp}, which show that Virgo provides the most stringent constraints among the galaxy clusters analysed. For Virgo, we adopt a NFW profile with scale radius $r_s=335.10 \, \rm{kpc}$, scale density $\rho_s=305646\, M_\odot\,\mathrm{kpc}^{-3}$~ \cite{DiMauro:2023qat}. The cluster is located at a distance $d_L =15.46 \, \rm{Mpc}$ and subtends and angle by the virial radius $\theta_{200}=6.32 \,\rm{deg}$~\cite{DiMauro:2023qat}.

We neglected the radial suppression factor \( f_{\rm tidal}(r) \) that accounts for tidal stripping of cusps in the inner regions of the host halo, given that its total annihilation signal is relatively modest, of order 5\%, and does not significantly affect the derived constraints~\cite{Crnogorcevic:2025nwp}.

By combining our estimate of \( \rho_{\rm eff, 0} \) from the enhanced power spectrum with the Virgo cluster profile, we compute the integrated $\rm{J}$ factors and derive constraints on the annihilation cross section as a function of the dark matter mass, by rescaling the results in Ref.~\cite{Crnogorcevic:2025nwp}.

\begin{figure}[h]
    \centering
    \includegraphics[width=1.0\linewidth]{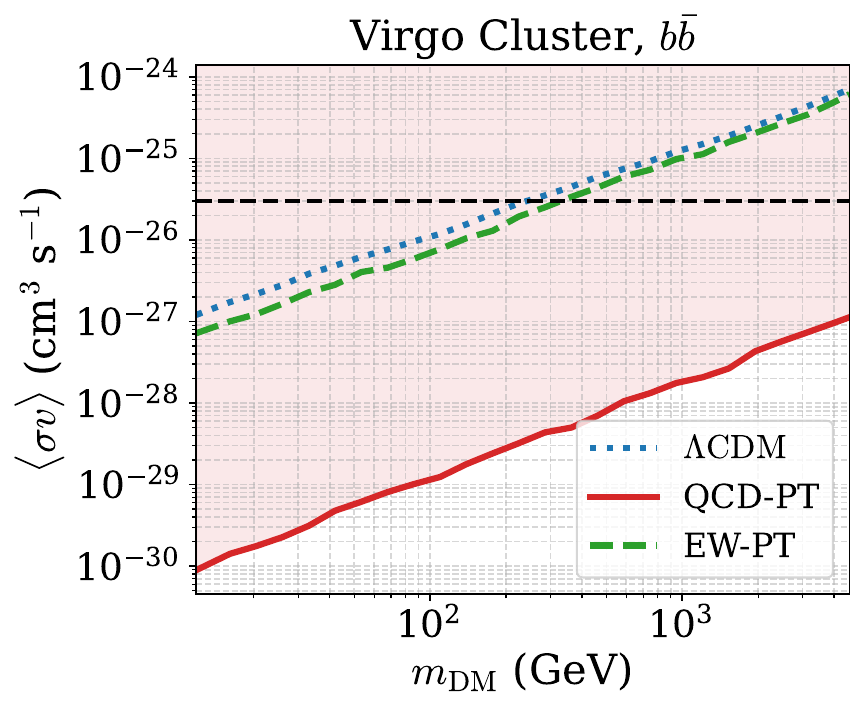}
    \caption{    Upper limits on the dark matter annihilation cross-section to a $b\bar{b}$ final state in the Virgo cluster, assuming a primordial magnetic field generated at the QCD phase transition (QCD-PT, red) or at the electroweak phase transition (EW-PT, green dashed). The $\Lambda$CDM baseline \htb{with self-annihilating dark matter} (blue dotted) and the canonical thermal cross-section (black dashed) are shown for comparison. The QCD-PT scenario yields the strongest bounds, ruling out thermal WIMPs up to multi-TeV masses. }
    \label{fig:PTs}
\end{figure}

\section{Results}
\label{sec:results}

\begin{figure}[t]
  \centering
  \includegraphics[width=\linewidth]{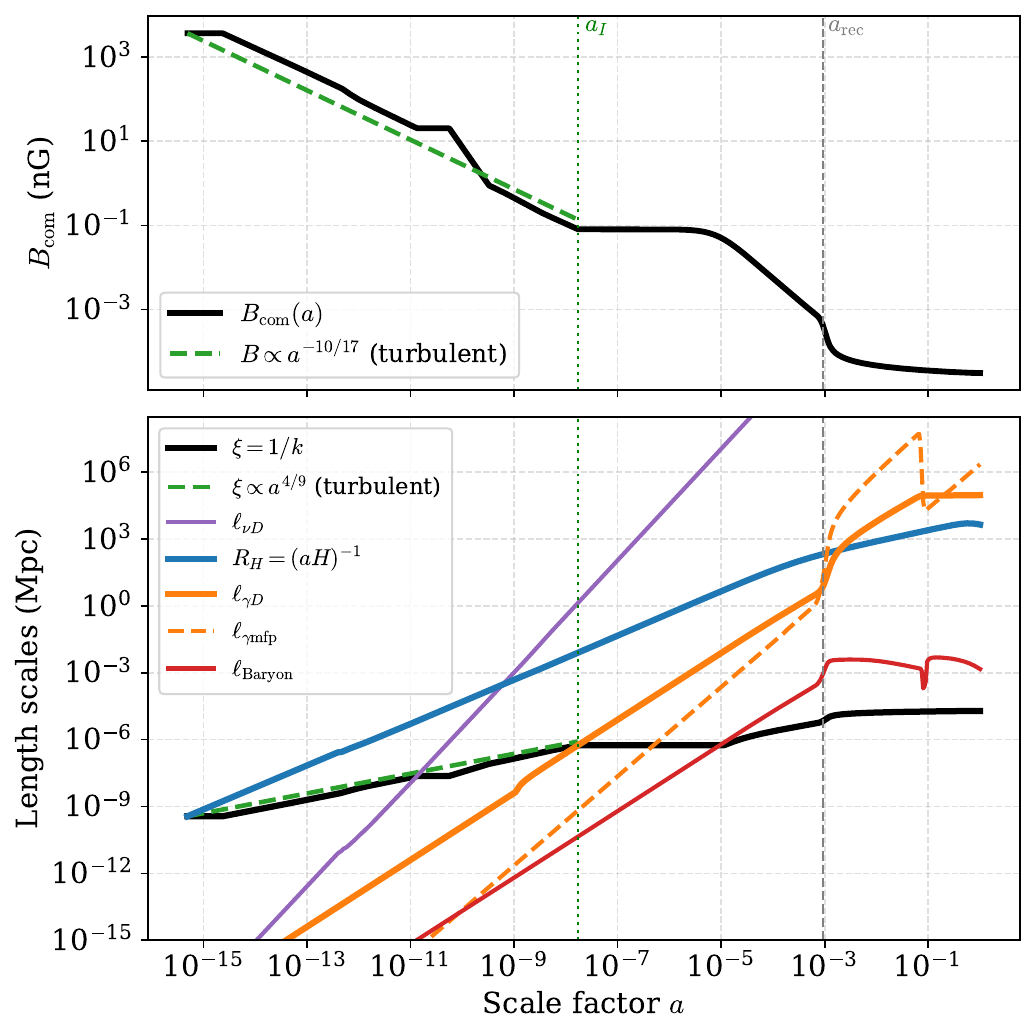}
  \caption{%
    \htb{Cosmological evolution of the comoving magnetic field and relevant comoving length scales for a field generated at an electroweak–scale phase transition with $T_{\rm PT}=160\,\mathrm{GeV}$. 
    \textbf{Top:} $B_{\rm com}(a)$ (solid) with a turbulent guide $B\propto a^{-10/17}$ (dashed). 
    \textbf{Bottom:} Coherence length $\xi=1/k$ (black) alongside neutrino diffusion $\ell_{\nu D}$ (purple), horizon $R_H=(aH)^{-1}$ (blue), photon diffusion $\ell_{\gamma D}$ (orange), photon mean free path $\ell_{\gamma\rm mfp}$ (orange, dashed), and the baryon thermal–pressure scale $\ell_{\rm Baryon}$ (red). 
    Vertical lines mark $a_I$ and $a_{\rm rec}$. 
    PMFs evolve under viscous damping when $\xi$ is below the relevant diffusion/free–streaming scales.
  }}
  \label{fig:magnetic_evolution}
\end{figure}

Here we discuss our results for the annihilation cross section constraints derived from the enhanced small-scale matter power spectrum induced by PMFs, as traced by the dark matter annihilation signal from the Virgo cluster. We focus on annihilation into the $b\bar{b}$ channel, since it yields the strongest constraints~\cite{Crnogorcevic:2025nwp}. \htb{Throughout this section we assume non-helical primordial magnetic fields evolving according to the description in Sec.~\ref{sec:PMFandHaloFor} — reconnection–driven turbulence for $Re>1$ and viscous damping for $Re<1$ — which is later illustrated in Fig.~\ref{fig:magnetic_evolution} for the case of the electroweak phase transition. This is the same evolution behaviour applied to all benchmark scenarios analysed in this work.} \htb{Note that helical magnetic fields, which evolve according to $B^2\xi=\mathrm{const}$~\cite{Durrer:2013pga}, could yield larger field strengths and coherence lengths than the non-helical case considered here, potentially leading to stronger constraints on dark matter annihilation.} 

Fig.~\ref{fig:virgoPlanck} shows the upper limits on the dark matter annihilation cross-section into $b\bar{b}$ final states in the Virgo cluster, assuming a primordial magnetic field with present-day strength $B_0 = 0.0034 \, \mathrm{nG}$, corresponding to the best-fit value from the combination of Planck CMB data and DESI Year~1 BAO measurements~\cite{Jedamzik:2025cax}. The red curve \htb{\htb{B$\Lambda$CDM} model, which extends $\Lambda$CDM by including both a primordial magnetic field and self-annihilating dark matter.} The magnetic field coherence length is set by  MHD turbulence, as mentioned above. 

Compared to the $\Lambda$CDM scenario \htb{with self-annihilating dark matter}  (blue dashed line) \htb{and kinetic decoupling temperature $T_{\rm kd}= 30 \, \rm{GeV}$}, adding a primordial magnetic field enhances the small-scale structure and leads to significantly stronger gamma-ray constraints. In this scenario, annihilation cross-sections are excluded up to six orders of magnitude below the thermal relic cross section (black dashed line), across the full dark matter mass range from below $10 \, \mathrm{GeV}$ to above $4 \, \mathrm{TeV}$. The improvement in the fit to Planck and DESI data between B$\Lambda$CDM and $\Lambda$CDM corresponds to $\Delta\chi^2 = -4.7$ (approximately a $2\sigma$ preference), and the resulting impact on indirect detection constraints is significant.

Fig.~\ref{fig:PTs} illustrates two benchmark scenarios for primordial magnetic fields generated at the QCD and electroweak phase transitions. For the electroweak-scale magnetogenesis case, we assume magnetic fields are produced at $T_{\mathrm{PT}} = 160\, \mathrm{GeV}$ with maximal energy density, i.e., $\rho_B = \rho_{\mathrm{SM}}$, and coherence length $\xi$ equal to the horizon size at that time, $\xi = (aH)^{-1}$. This choice represents an optimistic upper bound, as more realistic magnetogenesis scenarios are expected to yield smaller field strengths.

Fig.~\ref{fig:magnetic_evolution} shows the evolution of the magnetic field strength and coherence length after electroweak magnetogenesis, \htb{obtained with the prescription described in Sec.~\ref{sec:PMFandHaloFor}. This example illustrates the evolution law that we apply consistently to all benchmark scenarios explored in this work. The two distinct regimes are visible in different stages of the evolution}. The first regime corresponds to when the plasma kinetic Reynolds number ($Re$) is above unity, \htb{the coherence scale $\xi$ lies below both the photon and neutrino mean–free paths, and the field evolves turbulently. In this phase the inverse cascade conserves the Hosking integral ($B^4 \xi^5 = \text{const}$~\cite{Hosking:2022umv}), leading to the characteristic power–law evolution $B_{\rm com} \propto a^{-10/17}$ and $\xi \propto a^{4/9}$, consistent with the scaling expected for magnetically dominated turbulence~\cite{Banerjee:2004df}. The coherence scale is well approximated by} $\xi \sim \frac{0.1 V_A}{aH}$. \htb{The turbulent evolution ceases once $\xi$ overtakes the neutrino mean–free path, $\ell_{\nu \mathrm{mfp}}$, since neutrinos begin to diffuse efficiently and induce viscous drag on the plasma. The evolution resumes once the neutrino mean–free path grows beyond the horizon size, at which point neutrinos become fully free–streaming, decouple from the plasma, and can no longer damp the turbulence.}

\htb{As the universe expands further, the coherence scale eventually exceeds the photon diffusion length, $\ell_{\gamma D}$. At this point, photon viscosity becomes the dominant source of drag on the plasma, and the magnetic turbulence transitions into a photon–dominated viscous regime. In this phase, the eddy motions on scales below $\ell_{\gamma D}$ are damped, and the cascade slows down substantially, leading to a much weaker decay of the comoving field strength.}  

\htb{Here we note that the Boltzmann equations used to quantify the impact of PMFs on dark–matter perturbations are valid only once $\xi$ exceeds the photon mean–free path, $\ell_{\gamma \mathrm{mfp}}$. There is therefore a short transition interval between the end of the turbulent phase ($\xi \sim \ell_{\gamma D}$) and the onset of the well–defined viscous regime ($\xi \sim \ell_{\gamma \mathrm{mfp}}$). Although the microphysics in this regime is somewhat uncertain, this ambiguity has negligible impact on the final dark–matter perturbations, since the PMF–induced source terms are known to evolve toward an attractor solution~\cite{Ralegankar:2024arh}.}

Under these assumptions, magnetic field generation at the electroweak phase transition leads to a present-day comoving magnetic field strength of $B_0 \approx 3.06 \times 10^{-5}\, \mathrm{nG}$. As seen in Fig.~\ref{fig:PTs}, this field induces a modest enhancement of the $J$-factor. Consequently, the electroweak magnetogenesis scenario (green curve) excludes the canonical thermal relic cross section only for dark matter masses up to $m_{\chi} \lesssim 300\, \mathrm{GeV}$, resulting in constraints that are only marginally stronger than those \htb{obtained in \(\Lambda\)CDM with self-annihilating dark matter but} without primordial magnetic fields (blue dashed).   Future indirect detection constraints on the WIMP annihilation signal for higher mass WIMPs are expected from the Cerenkov Telescope Array \cite{CTA:2020qlo}.

For the QCD-scale magnetogenesis scenario, we assume that magnetic fields are generated at $T_{\mathrm{PT}} = 150\, \mathrm{MeV}$ with maximal energy density, $\rho_B = \rho_{\mathrm{SM}}$, and a coherence length equal to the horizon size at that time, $\xi = (aH)^{-1}$. Under the B\(\Lambda\)CDM hypothesis (red curve), this enhancement of small‐scale power increases the J-factor by roughly five orders of magnitude relative to \(\Lambda\)CDM \htb{with self-annihilating dark matter}, thereby excluding the canonical thermal‐relic cross section for dark matter masses from below \(10\ \mathrm{GeV}\) up to beyond \(4\ \mathrm{TeV}\).

These two benchmark scenarios demonstrate that even sub-nanogauss magnetic fields, naturally arising from cosmological phase transitions, can significantly enhance the small-scale matter power spectrum. This enhancement, in turn, leads to the early formation of dense minihalos with prompt cusps, and can result in a substantial boost to the dark matter annihilation rate.

\begin{figure}[H]
    \centering
    \includegraphics[width=0.9\linewidth]{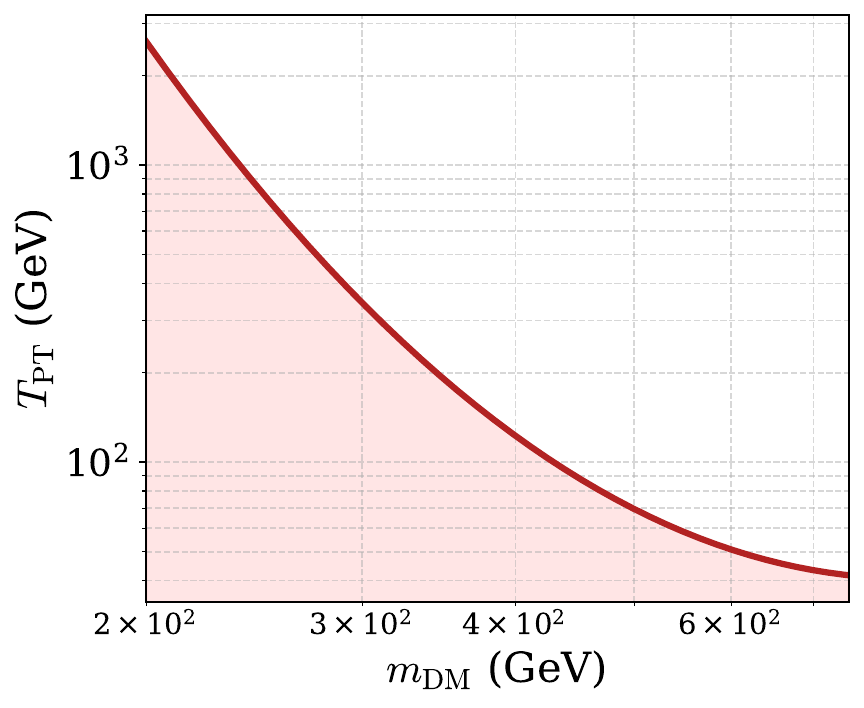}
    \caption{Lower bound on the temperature of the phase transition $T_{\rm PT}$ as a function of the dark matter mass $m_{\rm DM}$, assuming that the magnetic energy density at generation equals that of the Standard Model plasma. \htb{The evolution of the magnetic field strength and coherence length is given in Sec.~\ref{sec:PMFandHaloFor}.} For a given dark matter mass, values of $T_{\rm PT}$ below the curve are excluded under the assumption of thermal-relic annihilation.}
    \label{fig:Tpt_vs_mDMfig}
\end{figure}

To generalise beyond these benchmarks, Fig.~\ref{fig:Tpt_vs_mDMfig} shows the lower bound on the temperature of the phase transition, $T_{\rm PT}$, as a function of the dark matter mass, $m_{\rm DM}$. As in the benchmark scenarios discussed above, this analysis assumes that the primordial magnetic field is generated at a phase transition with energy density equal to that of the Standard Model plasma and a coherence length of the order of the horizon size. The resulting magnetic field strength is constrained by indirect detection limits, under the assumption of a thermal-relic annihilation cross section. For phase transitions occurring at \( T_{\rm PT} \lesssim 100 \, \mathrm{GeV} \), dark matter masses below \( 400 \, \mathrm{GeV} \) are excluded in this setup. This bound delineates the viable region for magnetogenesis from phase transitions in scenarios where dark matter is a thermal WIMP.

\begin{figure}[t]
  \centering
  \begin{subfigure}[b]{0.38\textwidth}
    \centering
    \includegraphics[width=\linewidth]{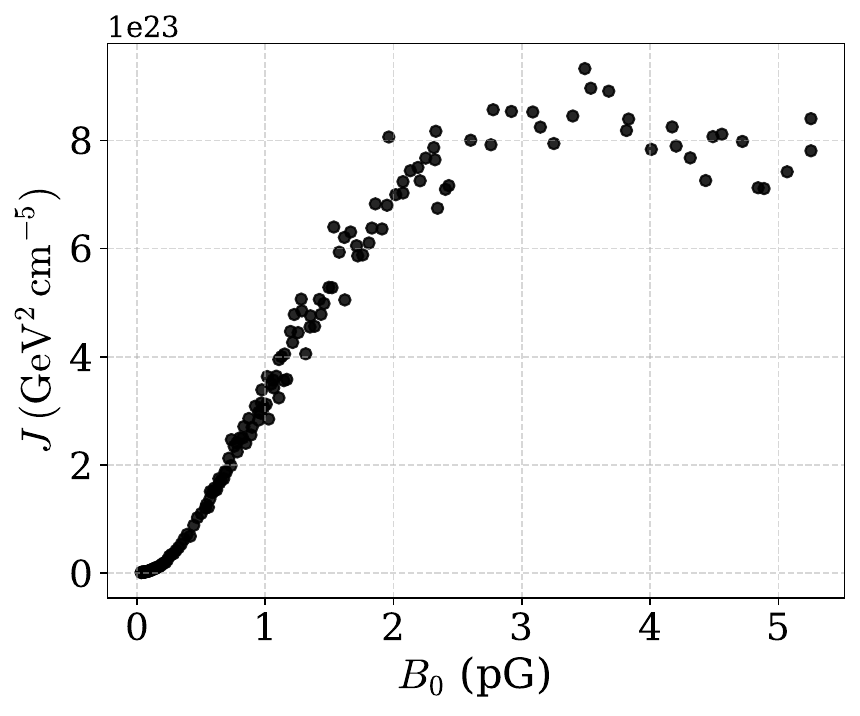}
    \caption{$J$ as a function of the present-day PMF amplitude $B_{0}$.}
    \label{fig:JvsB0}
  \end{subfigure}\hfill
  \begin{subfigure}[b]{0.38\textwidth}
    \centering
    \includegraphics[width=\linewidth]{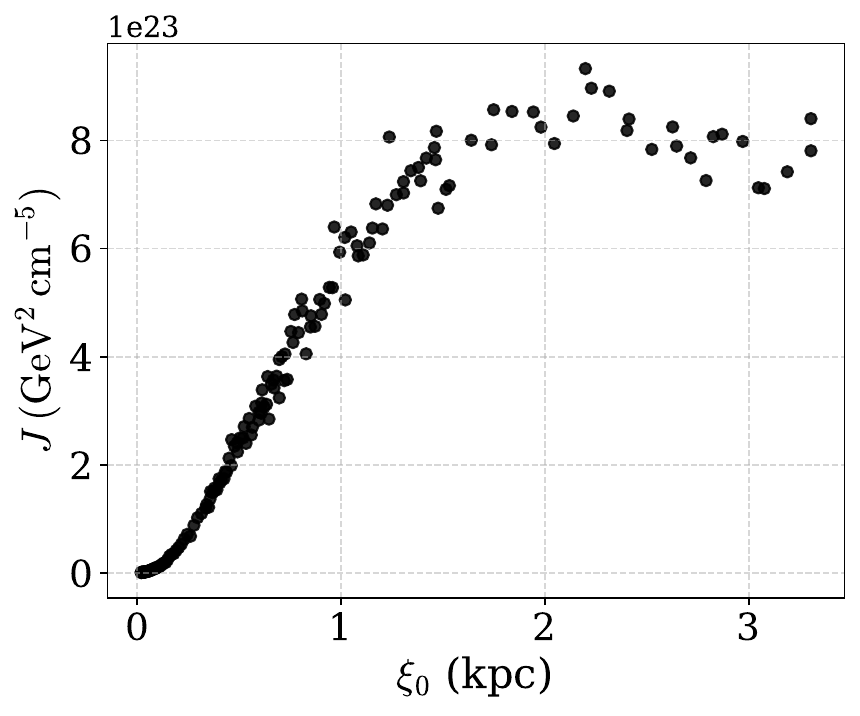}
    \caption{$J$ as a function of the present-day coherence length $\xi_{0}$.}
    \label{fig:JvsXi}
  \end{subfigure}
  \caption{%
    \htb{Scaling of the predicted annihilation $J$–factor with (a) the present-day PMF strength $B_{0}$ and (b) the present-day coherence length $\xi_{0}$.
    The results correspond to the Virgo cluster, assuming a dark-matter mass of $m_{\rm DM}=100~\mathrm{GeV}$ and a kinetic-decoupling temperature of $T_{\rm kd}=30~\mathrm{GeV}$.}}
  \label{fig:JvsB0Xi}
\end{figure}

\htb{The corresponding impact of the magnetic field parameters on the annihilation
flux is illustrated in Fig.~\ref{fig:JvsB0Xi}. The $J$-factor shows a clear
positive correlation with both the present-day magnetic field amplitude,
$B_0$, and the comoving coherence length, $\xi_{0}$. In each case,
$J$ rises steadily with increasing field strength or coherence, approximately
linearly at small values and flattening towards higher ones. This trend
highlights that stronger or more extended primordial fields lead to an overall
enhancement of the expected annihilation signal, consistent across the
parameter space explored. These plots permit to directly visualise the scaling of the annihilation flux with the magnetic field parameters, independently of any specific magnetogenesis scenario or phase transition assumption.
}

\section{Conclusions}

In this study, we have investigated how primordial magnetic fields (PMFs) enhance the small-scale matter power spectrum, leading to the early formation of dense dark matter minihalos with prompt central cusps. 
These cusps, characterised by steep inner density profiles scaling as $\rho \propto r^{-3/2}$, result in a significant enhancement of the dark matter annihilation rate due to their extreme central densities.

By modeling the abundance and structure of the prompt cusps using peak statistics and ellipsoidal collapse, and comparing the predicted annihilation signal to gamma-ray observations from the Virgo cluster, we have derived constraints on the annihilation cross section of self-annihilating dark matter. 
\htb{Our analysis considers three representative PMF scenarios: two benchmarks linked to cosmological phase transitions—specifically, the electroweak and QCD transitions—where we assume maximal magnetic energy density and horizon-sized coherence length at generation (upper-limit cases), and a data-driven scenario corresponding to the best-fit present-day PMF amplitude inferred from DESI BAO and Planck CMB measurements.}

\htb{The phase-transition benchmarks illustrate the largest plausible impact of primordial magnetogenesis, while the CMB–DESI case represents a realistic, observationally motivated level of intergalactic magnetisation.}
Across all cases, we find that the presence of PMFs substantially strengthens the indirect-detection limits on thermal WIMPs relative to standard $\Lambda$CDM.

\htb{Magnetic fields with amplitudes matching the DESI–Planck best-fit values are already in strong tension with self-annihilating WIMPs across a wide mass range extending beyond the TeV scale. 
In the phase-transition benchmarks, magnetic fields generated at the electroweak and QCD epochs would exclude thermal relics with masses below approximately $300\,\mathrm{GeV}$ and $3\,\mathrm{TeV}$, respectively. 
These phase-transition benchmarks serve as upper bounds on the possible impact of early magnetogenesis. 
More realistic phase-transition scenarios—where only a fraction of the plasma energy is transferred into magnetic fields or where the coherence length is sub-horizon—would naturally yield weaker quantitative limits.}  

\htb{Nevertheless, the qualitative conclusion remains robust: whenever primordial magnetic fields enhance the formation of early dark-matter structures, indirect-detection limits must be revisited, as standard $\Lambda$CDM assumptions no longer capture the correct small-scale dynamics.}

In addition, we have presented a complementary constraint based on the energy scale of the phase transition responsible for PMF generation. 
Assuming that the magnetic field is produced at a transition with an energy density equal to that of the Standard Model plasma and with a coherence length of the order of the horizon size, we compute the resulting magnetic field strength and determine the corresponding gamma-ray signal under the assumption of a thermal annihilation cross section. 
As shown in Fig.~\ref{fig:Tpt_vs_mDMfig}, for each value of the dark matter mass, temperatures below the exclusion curve are ruled out, since they would produce fields strong enough to generate prompt cusps that overproduce the annihilation signal. 
\htb{Relaxing the maximal-energy assumption would shift these curves toward smaller dark-matter masses, weakening the bounds quantitatively but leaving the overall trend unchanged.}

Our findings highlight the power of indirect detection not only to constrain the microscopic properties of dark matter but also to probe the early-universe processes that could have seeded cosmic magnetic fields. 
\htb{Even sub-nanogauss primordial fields can act as efficient amplifiers of annihilation signatures, linking cosmological magnetogenesis with observable indirect-detection signals and providing a new window into the interplay between particle physics and the magnetised early Universe.}

\section*{Acknowledgments}
MF and MOOR are supported by the STFC under UKRI grant ST/X000753/1.

\bibliography{references}
\bibliographystyle{apsrev4-1}

\end{document}